\documentclass[prb,twocolumn,amsmath,showpacs,floatfix]{revtex4}

\usepackage{graphicx}

\begin{document}

\title{
  Quantitative calculations of the excitonic energy spectra of
  semiconducting single-walled carbon nanotubes within a
  $\pi$-electron model 
}

\author{Zhendong Wang} 
\author{Hongbo Zhao} 
\author{Sumit Mazumdar}
\affiliation{Department of Physics, University of Arizona, Tucson,
  Arizona 85721, USA}

\date{\today} 

\begin{abstract}
  Using Coulomb correlation parameters appropriate for $\pi$-conjugated
  polymers (PCPs), and a nearest neighbor hopping integral that is
  arrived at by fitting the energy spectra of three zigzag
  semiconducting single-walled carbon nanotubes (S-SWCNTs), we are
  able to determine quantitatively the exciton energies and exciton
  binding energies of 29 S-SWCNTs within a semiempirical $\pi$-electron
  Hamiltonian that has been widely used for PCPs.  Our work
  establishes the existence of a deep and fundamental relationship
  between PCPs and S-SWCNTs.
\end{abstract}


\maketitle

\section{Introduction}

The photophysics of semiconducting single-walled carbon nanotubes
(S-SWCNTs) are of strong current interest because of their potential
applications in optoelectronics. \cite{Misewich03,Hertel00} Recent
theoretical works have emphasized the Coulomb-induced binding between
the optically excited electron and hole in these systems.
\cite{Ando97,Lin00,Kane,Spataru, Chang04, Zhao04, Perebeinos} The
excitonic energy spectra of S-SWCNTs are now understood qualitatively:
there occur in these nanotubes a series of energy manifolds labeled by
an index $n$ = 1, 2, ... etc., with each manifold containing both
optically allowed and dark exciton states, and a continuum band
separated from the optical exciton by a characteristic exciton binding
energy.  Nonlinear optical absorption measurements have demonstrated
that distinct energy gaps occur between the lowest two-photon allowed
states and the $n$ = 1 optical excitons.
\cite{Wang05,Dukovic05,Maultzsch05,Zhao06} This energy gap is a lower
bound to the binding energy of the $n$ = 1 optical
exciton.\cite{Zhao06} Nonlinear absorption measurements involving the
$n=1$ continuum band \cite{Ma05} and the $n=2$ exciton
\cite{Rothberg06} have also been performed.

Exciton formation in S-SWCNTs is a consequence of direct Coulomb
electron-electron (e-e) interaction, taking proper account of which is
a notoriously difficult many-body problem. Theoretical calculations of
excitons and continuum band energies in the S-SWCNTs are therefore
necessarily approximate.  One semiquantitative approach that has been
popular employs an {\it ab initio} approach for the ground state,
which is followed by the determination of the quasiparticle energies
within the GW approximation and the solution of the Bethe-Salpeter
equation of the two-particle Green's function. \cite{Spataru,Chang04}
This approach becomes difficult to implement for chiral S-SWCNTs with
large diameters and large unit cells, precisely the systems for which
the nonlinear optical measurements are being performed.
\cite{Wang05,Dukovic05,Maultzsch05,Zhao06,Ma05,Rothberg06} Scaling
relationships for the exciton binding energies in wider chiral
nanotubes have therefore been proposed.  \cite{Perebeinos,Kane}
Understanding nonlinear absorption within such indirect approaches is
difficult.

An alternate approach to the exciton problem in the S-SWCNTs is to
adopt a semiempirical $\pi$-electron approximation \cite{Pariser} that
is widely used to describe {\it planar} $\pi$-conjugated systems.  The
model assumes that the lowest energy excitations in planar conjugated
systems involves the $\pi$-electrons only and ignores the electrons
occupying orthogonal $\sigma$-bands.  Since $\pi$--$\pi^*$ excitation
energies decrease rapidly with increasing system size, while the
$\sigma$ and $\sigma^*$ bands are nearly dispersionless, the
approximation is excellent for large systems. Thus $\pi$-conjugated
polymers (PCPs) have been discussed extensively within the
semiempirical Pariser-Parr-Pople (PPP) Hamiltonian, \cite{Pariser}
\begin{equation}
\begin{split}
\label{ppp}
H = & -\sum_{\langle ij \rangle,\sigma}t_{ij}
         (c_{i\sigma}^\dag c_{j\sigma} + c_{j\sigma}^\dag c_{i\sigma}) 
      + U \sum_i n_{i\uparrow}n_{i\downarrow} \\
    & + \frac{1}{2} \sum_{i\neq j} V_{ij}(n_i-1)(n_j-1) .
\end{split}
\end{equation}
Here $c^\dagger_{i \sigma}$ creates a $\pi$-electron with spin
$\sigma$ ($\uparrow$, $\downarrow$) in the $p_z$ orbital of the $i$th
carbon atom, $\langle ij\rangle$ implies nearest neighbor (n.n.)
atoms $i$ and $j$, $n_{i \sigma} = c^\dagger_{i \sigma}c_{i \sigma}$
is the number of $\pi$-electrons with spin $\sigma$ on the atom $i$,
and $n_i = \sum_{\sigma}$ is the total number of $\pi$-electrons on
the atom. The parameter $t_{ij}$ is the one-electron hopping integral
between $p_z$ orbitals of n.n.\ carbon atoms, $U$ is the on-site e-e
repulsion between two $\pi$-electrons occupying the same carbon atom
$p_z$ orbital, and $V_{ij}$ is the intersite e-e interaction.

The semiempirical model suffers from the disadvantage that
determination of the parameters is commonly difficult.  On the other
hand, many-body problems that would be formidable within the {\it ab
  initio} approach, such as the enhancement of the ground state
bond-alternation in polyacetylene by e-e interactions,
\cite{Baeriswyl92} or the occurrence of the lowest two-photon state
below the optical state in finite polyenes and polyacetylene,
\cite{Soos93} can be understood within Eq.~(1). The semiempirical
approach should thus be taken as complementary to the {\it ab initio}
one.

The above observations were the basis of our calculations within
Eq.~(1) of the electronic structures and exciton binding energies,
\cite{Zhao04} and more recently, nonlinear optical absorptions
\cite{Zhao06} in a limited number of S-SWCNTs, in spite of their
nonplanarity. Justification of this procedure can also be found in
earlier work claiming that curvature effects of fullerene molecules
can be included within Eq.~(1) by proper modifications of the
parameters. \cite{Haddon93} The e-e interaction and hopping integral
parameters used in our calculations were the same ones that had been
used earlier for the PCP poly-paraphenylenevinylene (PPV).
\cite{Chandross97} The calculated energy differences ($\sim$ 0.3--0.4
eV) between dominant two-photon states observed in ultrafast
spectroscopy and the $n=1$ optical excitons, in S-SWCNTs with
diameters d $\geq$ 0.8 nm, were remarkably close to what are observed
experimentally. \cite{Zhao06} Other experimental work
\cite{Wang05,Dukovic05,Maultzsch05,Ma05} have also determined the
binding energy of the $n = 1$ optical exciton in S-SWCNTs with
diameters 0.8--1.0 nm to be $\sim$0.4 eV, in agreement with
Ref.~\onlinecite{Zhao04}.  Very recent work has also confirmed the
predicted binding energy of the $n = 2$ exciton in two different
S-SWCNTs.\cite{Rothberg06}

The calculated {\it absolute energies} of the $n$ = 1 and 2 excitons
S-SWCNTs in our earlier work, \cite{Zhao04,Zhao06} however, are much
larger than the experimental values (by as much as 0.5 eV). Equally
importantly, comparisons of experimental and calculated exciton
binding energies were based on considerations of diameters alone:
while the nanotubes probed experimentally are chiral with large
diameters, the theoretical calculations were either for zigzag
nanotubes or for narrow chiral nanotubes, both with smaller unit
cells.  Direct comparisons of theory and experiment for the same
systems were thus not possible. Finally, because the theoretical
calculations could be performed for only a handful of nanotubes,
family relationships, established experimentally, \cite{Bachilo02}
could not be demonstrated within the theoretical model.

In the present paper, we report calculated energy spectra of 29
S-SWCNTs, only 9 of which are zigzag, within Eq.~(1). The diameters of
the S-SWCNTs we consider range from 0.56 to 1.51 nm. These
calculations have been possible because of substantive improvements in
our computational techniques, while the improved results are a
consequence of careful parametrization of the n.n.\ hopping integral
$t_{ij} =t$ that takes into account curvature effects (see below).
For each S-SWCNT we calculate the absolute energies $E_{11}$ and
$E_{22}$ of the $n$ = 1 and 2 optical excitons, and their binding
energies $E_{b1}$ and $E_{b2}$, respectively.  We compare all
theoretical quantities to experimentally determined ones.
\cite{Wang05,Dukovic05,Maultzsch05,Zhao06,Ma05,Rothberg06,Bachilo02,Weisman03,Fantini04}
The large number of S-SWCNTs that could be considered allows us to
investigate family relationships based on calculations of the energy
ratio $E_{22}/E_{11}$ and to compare these data to experiments.
\cite{Bachilo02} We find excellent agreement between the theory and
experiments. Our work demonstrates convincingly that the photophysics
of S-SWCNTs and PCPs can be understood within the same general
theoretical framework, albeit with different hopping integrals.

\section{Method and parametrization of $\pi$-electron model}

We use the single configuration interaction (SCI) approximation to
compute the energies of the one electron-one hole excitations of the
Hamiltonian in Eq.~(1). The SCI approximation includes configuration
mixings only between excitations that are singly excited from the
Hartree-Fock (HF) ground state.  The justifications for the
approximation have been given before.\cite{Zhao04,Zhao06}

We discuss next how the parametrizations within Eq.~(1) were reached.
Since e-e interactions depend only on the distance and not on the
curvature of the S-SWCNTs, it is logical to parametrize the $V_{ij}$
exactly as in the PCPs, \cite{Chandross97} viz.,
\begin{equation}
\label{parameters}
V_{ij}=\frac{U}{\kappa\sqrt{1+0.6117 R_{ij}^2}}, 
\end{equation}
where $R_{ij}$ is the distance between carbon atoms $i$ and $j$ in
\AA.  The two free parameters in Eq.~(\ref{parameters}) are $U$ and
$\kappa$; the latter is introduced to take into account the dielectric
screening due to the medium. \cite{Chandross97} $U$ = 11.26 eV and
$\kappa$ = 1 correspond to the standard Ohno parametrization for
finite molecules.  \cite{Ohno} SCI calculations of the optical
absorption in PPV with five different $U$ (between 0 to 10 eV) and
three different $\kappa$ (1, 2, and 3) had indicated that only with
$U$ = 8 eV and $\kappa$ = 2 was it possible to fit all four absorption
bands at 2.4, 3.7, 4.7, and 6.0 eV in PPV.  The same $U$ and $\kappa$
were then used for quantitative calculations of nonlinear and triplet
absorptions in PPV. Experimentally determined energies of the
two-photon state that dominates nonlinear optical spectroscopy
\cite{Frolov00} and the lowest triplet state \cite{Monkman99} agreed
remarkably well with the theory.  We have used the same $U$ and
$\kappa$ for S-SWCNTs.

\begin{table}
\caption{
  Calculated and experimental (Ref.~\onlinecite{Weisman03}) 
  $n$ = 1 and 2 exciton energies 
  for three zigzag S-SWCNTs. 
}
\begin{ruledtabular}
\begin{tabular}{cccccc}
 & & \multicolumn{2}{c}{$E_{11}$ (eV)} 
   & \multicolumn{2}{c}{$E_{22}$ (eV)} \\
 ($n$,$m$) & $t$ (eV) & SCI & Expt. & SCI &  Expt. \\ \hline
(10,0) & 1.8 & 1.10 & 1.07 & 1.97 & 2.31 \\
       & 1.9 & 1.14 &      & 2.05 &      \\
       & 2.0 & 1.18 &      & 2.13 &      \\
       & 2.4 & 1.33 &      & 2.45 &      \\ 
(13,0) & 1.8 & 0.90 & 0.90 & 1.59 & 1.83 \\
       & 1.9 & 0.93 &      & 1.65 &      \\
       & 2.0 & 0.96 &      & 1.71 &      \\
       & 2.4 & 1.08 &      & 1.96 &      \\ 
(17,0) & 1.8 & 0.73 & 0.80 & 1.24 & 1.26 \\
       & 1.9 & 0.75 &      & 1.28 &      \\
       & 2.0 & 0.77 &      & 1.32 &      \\
       & 2.4 & 0.87 &      & 1.50 &    
\end{tabular}
\end{ruledtabular}
\end{table}

In the absence of any known procedure for determining $t$ in S-SWCNTs,
we had earlier \cite{Zhao04,Zhao06} chosen the value 2.4 eV that is
used for planar aromatic systems. \cite{Salembook} Curvature in the
S-SWCNTs implies smaller $\pi$--$\pi$ overlap between n.n.\ C atoms
and hence a smaller $t$.  We arrive at the proper $t$ by fitting the
calculated $E_{11}$ and $E_{22}$ for three different zigzag nanotubes,
(10,0), (13,0) and (17,0), against the corresponding experimental
quantities (see below for discussions of how the experimental
quantities were arrived at by previous authors).  The theoretical
exciton energies are for $U$ = 8 eV and $\kappa$ = 2, and four
different $t$ = 1.8, 1.9, 2.0, and 2.4 eV.  Our results are summarized
in Table~I, which clearly indicates that $t = 2.4$ eV is too large and
considerably better fits are obtained with $t$ = 1.8--2.0 eV. We have
chosen $t = 2.0$ eV for the complete set of calculations we report
below. The fits in Table~I improve with increasing nanotube diameter,
implying that strictly speaking the hopping integral is
diameter-dependent even within the range of diameters we consider.  We
do not attempt further fine tuning of parameters as this would
necessarily lead to loss of simplicity and generality.

As in previous work,\cite{Zhao04,Zhao06} we use open boundary
condition (OBC). Surface states due to dangling bonds at the nanotube
ends appear in the HF band structure, and are discarded at the SCI
stage of our calculations. The chiral S-SWCNTs we investigate have
gigantic unit cells. The number of unit cells we retain depend both on
the size of the unit cell and the convergence behavior of $E_{11}$.
The procedure involved calculating the standard tight-binding (TB)
band-structure with periodic boundary condition (PBC), and then
comparing the PBC $E_{11}$ with that obtained using OBC with a small
number of unit cells. The number of unit cells in the OBC calculation
is now progressively increased until the difference in the computed
$E_{11}$ between OBC and PBC is less than 0.004 eV (worst case).  It
is with this system size that the SCI calculations are now performed
using OBC.  Thus for example, our calculations for (7,0), (6,4), and
(7,5) SWCNTs are for 70, 16, and 5 unit cells, respectively,
containing 1960, 2432, and 2180 carbon atoms, respectively.  Since
energy convergences are faster in the calculations with nonzero e-e
interactions than the calculations in the TB limit, we are confident
that this procedure gives accurate results.  We retain an active space
of 100 valence and conduction band states each in the SCI
calculations.  Stringent convergence tests involving gradual increase
in the size of the active space indicate that the computational errors
due to the energy cutoff is less than 0.005 eV (worst case). In
addition to $E_{11}$ and $E_{22}$ we also calculate the corresponding
exciton binding energies $E_{b1}$ and $E_{b2}$, with the $n$ = 1 and 2
continuum band threshold energies defined within the SCI to be the
corresponding HF gaps.

\section{Results and discussion}

\begin{table*}
\caption{
  Comparison of calculated and experimental/empirical $n = 1$ and 2 exciton
  energies and binding energies. 
  The empirical exciton energies (Ref.~\onlinecite{Weisman03}) 
  and exciton binding energies (Ref.~\onlinecite{Dukovic05}) 
  are in parentheses.
}
\begin{ruledtabular}
\begin{tabular}{cccccccccc}
 & & \multicolumn{2}{c}{$E_{11}$ (eV)} 
   & \multicolumn{2}{c}{$E_{22}$ (eV)} 
   & \multicolumn{2}{c}{$E_{b1}$ (eV)} 
   & \multicolumn{2}{c}{$E_{b2}$ (eV)} \\
 ($n,m$)
 & $d$ (nm) 
 & SCI & Expt.\footnotemark[1] 
 & SCI & Expt. 
 & SCI & Expt.\footnotemark[3] 
 & SCI & Expt.\footnotemark[4]
\\ \hline
(7,0) &0.56 &1.58 &(1.29) &2.92 &(3.14)             &0.56 &(0.61) &0.79 \\
(6,2) &0.57 &1.55 &(1.39) &2.82 &(2.96)             &0.55 &(0.59) &0.72 \\
(8,0) &0.64 &1.44 &(1.60) &2.38 &(1.88)             &0.56 &(0.54) &0.57 \\
(7,2) &0.65 &1.41 &(1.55) &2.36 &(1.98)             &0.54 &(0.52) &0.56 \\
(8,1) &0.68 &1.34 &(1.19) &2.45 &(2.63)             &0.48 &(0.50) &0.65 \\
(6,4) &0.69 &1.33 & 1.42  &2.27 &2.13$^{a,b}$       &0.50 &(0.49) &0.56 \\
(6,5) &0.76 &1.24 & 1.27  &2.15 &2.19$^{a,b}$       &0.45 & 0.43  &0.54 \\
(9,1) &0.76 &1.24 & 1.36  &2.08 &1.79$^{a,b}$       &0.47 &(0.45) &0.51 \\
(8,3) &0.78 &1.21 & 1.30  &2.05 &1.87$^a$           &0.45 & 0.42  &0.50 \\
(10,0)&0.79 &1.18 &(1.07) &2.13 &2.26$^b$           &0.42 &(0.43) &0.57 \\
(9,2) &0.81 &1.17 &(1.09) &2.10 &2.24$^b$           &0.42 &(0.42) &0.55 \\
(7,5) &0.83 &1.15 & 1.21  &1.97 &1.93$^{a,b}$       &0.43 & 0.39  &0.49 &0.62$\pm$0.05\\
(8,4) &0.84 &1.13 & 1.11  &2.00 &2.11$^{a,b}$       &0.41 &(0.40) &0.51 \\
(11,0)&0.87 &1.11 &(1.20) &1.86 &(1.67)             &0.42 &(0.39) &0.46 \\
(10,2)&0.88 &1.09 &(1.18) &1.84 &1.68$^b$           &0.40 & 0.34  &0.45 \\
(7,6) &0.90 &1.08 & 1.11  &1.88 &1.92$^{a,b}$       &0.39 & 0.35  &0.47 \\
(9,4) &0.92 &1.06 & 1.13  &1.81 &1.72$^a$, 2.03$^b$ &0.39 & 0.34  &0.44 \\
(11,1)&0.92 &1.05 &(0.98) &1.89 &(2.03), 1.72$^b$   &0.37 &(0.37) &0.50 \\
(10,3)&0.94 &1.03 & 0.99  &1.84 &1.96$^{a,b}$       &0.37 &(0.36) &0.48 &0.49$\pm$0.05\\
(8,6) &0.97 &1.01 & 1.06  &1.75 &1.73$^{a,b}$       &0.37 & 0.35  &0.44 \\
(13,0)&1.03 &0.96 &(0.90) &1.71 &(1.83)             &0.34 &(0.33) &0.45 \\
(12,2)&1.04 &0.95 & 0.90  &1.69 &1.81$^a$           &0.33 &(0.33) &0.44 \\
(10,5)&1.05 &0.94 & 0.99  &1.62 &1.58$^{a,b}$       &0.35 &(0.32) &0.40 \\
(14,0)&1.11 &0.91 &(0.96) &1.54 &(1.44)             &0.34 &(0.31) &0.38 \\
(12,4)&1.15 &0.88 & 0.92  &1.51 &1.45$^a$           &0.32 & 0.27  &0.37 \\
(16,0)&1.27 &0.81 &(0.76) &1.44 &(1.52)             &0.28 &(0.27) &0.37 \\
(17,0)&1.35 &0.77 &(0.80) &1.32 &(1.26)             &0.28 &(0.25) &0.32 \\
(15,5)&1.43 &0.73 &(0.71) &1.29 &(1.35)             &0.25 &(0.24) &0.32 \\
(19,0)&1.51 &0.70 &(0.66) &1.24 &(1.30)             &0.24 &(0.23) &0.31 \\
\end{tabular}
\end{ruledtabular}
\footnotetext[1]{From Ref.~\onlinecite{Bachilo02}.}
\footnotetext[2]{From Ref.~\onlinecite{Fantini04}.}
\footnotetext[3]{From Ref.~\onlinecite{Dukovic05}.} 
\footnotetext[4]{From Ref.~\onlinecite{Rothberg06}.}
\end{table*}

In Table~II we have listed our calculated $E_{11}$, $E_{22}$,
$E_{b1}$, and $E_{b2}$ for 29 S-SWCNTs.  We compare each of these
quantities to those obtained by experimental investigators.
\cite{Bachilo02,Fantini04,Weisman03,Ma05,Dukovic05,Rothberg06} Nearly
half the exciton energies listed in Table~II as experimental were
obtained directly from spectroflurometric measurements\cite{Bachilo02}
or from resonant Raman spectroscopy.\cite{Fantini04} Using the
experimental data in Ref.~\onlinecite{Bachilo02}, Weisman and
Bachilo\cite{Weisman03} derived empirical equations for the exciton
energies of nanotubes for which direct experimental information do not
exist.  The experimental $E_{11}$ and $E_{22}$ in Table~I are obtained
from these empirical equations.  Dukovic {\it et al.}\cite{Dukovic05}
have given an empirical equation for the binding energy of the $n = 1$
exciton, which was also derived by fitting the set of $E_{b1}$
obtained from direct measurements.  Only two measured values exist
currently for the binding energy of $n = 2$ exciton. \cite{Rothberg06}
We make distinctions between the experimental and empirical data in
Table~II, but in the text below we refer to both as experimental
quantities.  In Fig.~1 we have plotted the theoretical and
experimental $E_{11}$ and $E_{22}$ against $1/d$, while in Fig.~1 inset
we show the errors in our calculations, $\Delta E_{11}$ and $\Delta
E_{22}$, defined as the calculated energies minus the experimental
quantities.  As seen in the figure the theoretical plots are nearly
independent of chirality, and depend primarily on diameter. This is a
consequence of Eq.~(1), within which the energetics depend only on the
conductivity.  We find excellent agreement between calculated and
experimental $E_{11}$ for $d > 0.75$ nm, with $|\Delta E_{11}| < 0.1$
eV. (The black arrow on the x-axis in Fig.~1 indicates $d=0.75$
nm.) The agreement for $d>1$ nm is even better with $|\Delta E_{11}| <
0.05$ eV.  The larger disagreement with experiment (and the greater
chirality dependency of the experimental energies) for $d<0.75$ nm is
due to the breakdown of the $\pi$-electron approximation.  The
disagreements between calculated and experimental $E_{22}$ are larger,
but even here the magnitude of the maximum error for $d > 0.75$ nm is
within 0.2 eV, which is the C--C bond stretching frequency that can
influence experimental estimation of exciton energies.
\cite{Perebeinos} The origin of the larger disagreement in the $n = 2$
region is the SCI approximation and not the $\pi$-electron model:
inclusion of higher order CI is more important in general for higher
energy states.

\begin{figure*}  
 \centering
 \includegraphics[clip,width=5in]{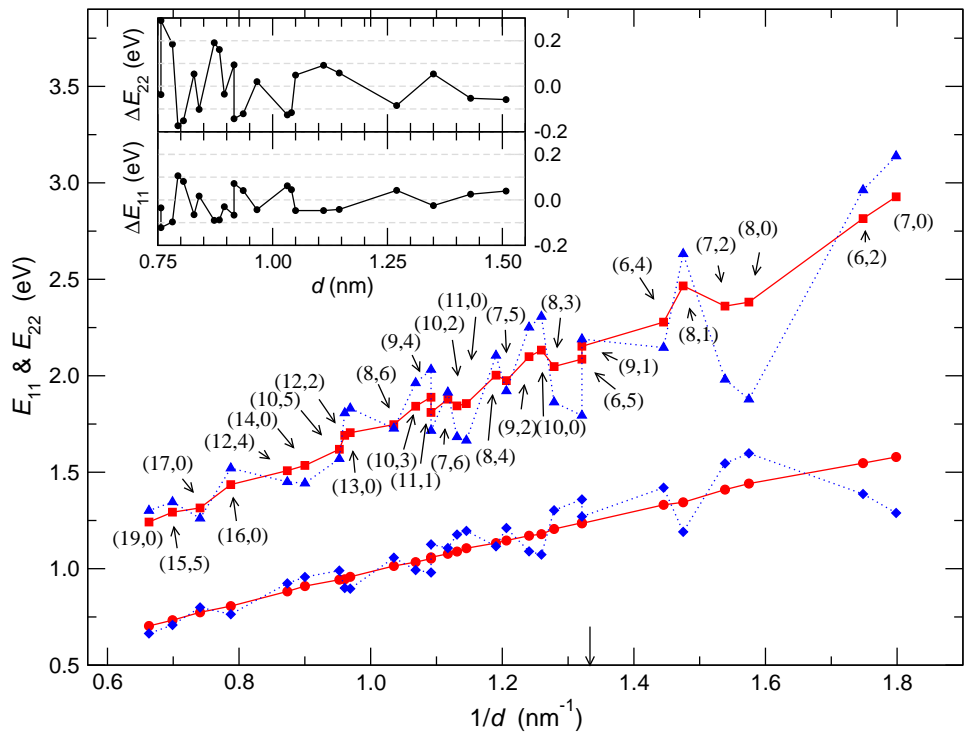}
 \caption{
   (Color online) Calculated (red solid line, circle and square symbols) 
       vs.\ experimental (blue dotted line, diamond and triangle symbols) 
       $E_{11}$ and $E_{22}$ for 29 S-SWCNTs.
       The inset shows errors $\Delta E_{ii}$ ($i$ = 1, 2) in the calculations,
       defined as the calculated minus the experimental or empirical
       energies.
 }
\end{figure*}

The agreement between the calculated and the experimental exciton
binding energies in Table~II is even more striking than the fits to
the absolute energies. The discrepancies between theory and experiment
is less than 10\%, which is the uncertainty in the empirical binding
energies. \cite{Dukovic05} Both $E_{b1}$ and $E_{b2}$ are
inversely proportional to the diameter $d$ and can be fitted
approximately by
\begin{subequations}
 \begin{eqnarray}
   E_{b1} \simeq &\frac{0.35}{d} \text{ eV}, \\ 
   E_{b2} \simeq &\frac{0.42}{d} \text{ eV}.
 \end{eqnarray}
\end{subequations}
The exciton binding energies depend weakly on chirality.  For both
binding energies, the fits of Eq.~(3) become better for larger
diameter nanotubes.  Eq.~(3a) is remarkably close to the empirical
formula $E_{b1}\simeq \frac{0.34}{d}$ eV given in previous
experimental work.  \cite{Dukovic05} Although there are not enough
experimental data to verify the $E_{b2}$ relation in Eq.~(3b), the
linear dependence against $1/d$ should be true for both binding
energies.  The calculated $E_{b1}$ and $E_{b2}$ are very close to
those obtained earlier by us for the wide nanotubes with $t = 2.4$ eV,
\cite{Zhao04,Zhao06} even as the calculated $E_{11}$ and $E_{22}$ are
now quite different.  This is simply a consequence of localization of
the electrons at the large $U/t$ considered here: exciton binding
energies in this case depend primarily on electron correlation.
Recall, for example, that in the limit where the intersite Coulomb
interaction is cut off beyond the n.n.\ interaction $V_1$, the exciton
binding energy is determined almost entirely by the difference in
Coulomb interactions $U-V_1$, \cite{Baeriswyl92} with $t$ playing a
weak role for $U/t \geq 4$.

\begin{figure}
 \centering
 \includegraphics[clip,width=3.375in]{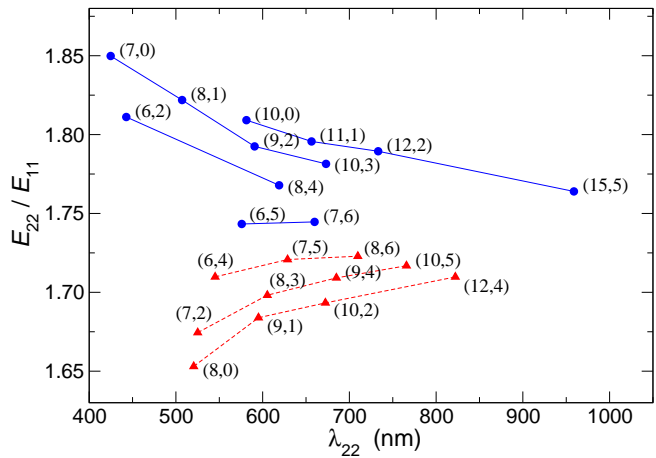}
 \caption{
   (Color online)
   Calculated $E_{22}$/$E_{11}$ vs.\ the 
   wavelength $\lambda_{22}$ corresponding to $E_{22}$.  
   The blue solid line with circle symbols 
   (red dashed line with triangle symbols)
   denote families $(n - m) \mod 3 = 1$ (2).
 } 
\end{figure}

Our ability to calculate $E_{11}$ and $E_{22}$ for a large number of
nanotubes allows us to demonstrate family behavior within Eq.~(1),
which was not possible before. It has been shown that in plots of
experimental $E_{22}/E_{11}$ against the wavelength $\lambda_{22}$
corresponding to $E_{22}$, families of ($n,m$) S-SWCNTs with the same
$n-m$ lie on the same continuous curves. \cite{Bachilo02} The
experimental energy ratios diverge from a central limiting number
$\sim$1.75, with the deviations increasing with $n-m$, and with the
direction of the deviation being positive for the families $(n - m)
\mod$ 3 = 1 and negative for the families $(n-m) \mod$ 3 = 2,
respectively. \cite{Bachilo02} In Fig.~2 we have plotted our
calculated $E_{22}/E_{11}$ against the calculated $\lambda_{22}$.
The calculated ratios do not show the same smooth behavior as in Ref.
\onlinecite{Bachilo02}, since as already noticed in Fig.~1 inset our
errors with $E_{22}$ are larger than those with $E_{11}$.
Nevertheless, {\it two different classes of behavior}, for the
families $(n - m) \mod$ 3 = 1 and $(n - m) \mod$ 3 = 2 are very clear.
The difference between the largest and the smallest ratios in the
narrow diameter (small $\lambda_{22}$) end is smaller than what is
seen experimentally, but this is once again merely a consequence of
the larger errors in our calculations for the narrower nanotubes.
More importantly, the limiting $E_{22}/E_{11}$ in the wide diameter
(large $\lambda_{22}$) end in Fig.~2 is very close to the experimental
limiting ratio of 1.75.

\section{Conclusion}

In summary, direct calculations of the energy spectra of a large
number of S-SWCNTs within the $\pi$-electron PPP model Hamiltonian and
with a fixed set of parameters show excellent agreement with
experiments. The magnitudes of two of the three free parameters, $U$
and $\kappa$ for Coulomb interactions, are the same as those used
earlier to calculate the energy spectrum of PPV;\cite{Chandross97} the
third free parameter $t$ is obtained by fitting the energetics of
three zigzag S-SWCNTs to account for the curvature effects. The
calculated exciton energies and exciton binding energies in the first
two manifolds are in excellent agreement with experiments, and the
``family behavior'' is demonstrated.  In the diameter range of our
calculated S-SWCNTs, we found linear dependence against $1/d$ for the
binding energies of excitons in both manifolds.  It is unlikely that
the agreement between theory and experiments is fortuitous.  The
Coulomb parameters used here were obtained after extensive search
through a very large parameter space for PPV, and the same parameters
successfully reproduced the energies of the dominant two-photon state
\cite{Frolov00} and the triplet energy spectrum \cite{Monkman99} of
PPV.  Similarly, slightly smaller hopping in S-SWCNTs than in PPV is
to be expected.\cite{Haddon93} Our work demonstrates a universality in
the photophysics of S-SWCNTs and PCPs that arises from their common
quasi-one-dimensionality and $\pi$-conjugation. The existing rich
literature on the photo- and device physics of PCPs can therefore
provide valuable guidance in the search for optoelectronic
applications of S-SWCNTs.

\begin{acknowledgments}
This research was supported by  NSF-DMR-0406604.
\end{acknowledgments}

\end{document}